\documentclass[11pt,a4paper]{article}
\usepackage[english]{babel}
\usepackage[utf8]{inputenc}
\usepackage{amsmath,amsthm,amsfonts,amssymb}
\usepackage{graphicx}
\usepackage{hyperref}
\usepackage{cite,float}
\usepackage{multicol}
\usepackage{authblk}

\usepackage{color}
\definecolor{cinza}{gray}{0.4}
\definecolor{verde}{rgb}{0,0.5,0}

\hypersetup{colorlinks=true, linkcolor=blue, citecolor=blue, filecolor=blue, urlcolor=blue}

\usepackage{caption}
\usepackage[margin=1.2in]{geometry}
\usepackage{verbatim}

\makeatletter
\newcommand{\xRightarrow}[2][]{\ext@arrow 0359\Rightarrowfill@{#1}{#2}}
\makeatother
\usepackage{booktabs}
\usepackage[svgnames,table]{xcolor}
\usepackage[tableposition=above]{caption}
\usepackage{pifont}

\usepackage[labelfont=bf,font=scriptsize]{caption}

\usepackage{tocloft}

\selectfont

\usepackage{empheq}

\title{Bohmian Quantization of a Nonminimal Coupling Cosmology}

\author[1]{Isaac Torres\footnote{its@ufpa.br}}
\author[2]{Felipe de Melo Santos}
\author[3]{Anderson Almeida da Piedade}

\affil[1]{\small Faculdade de F\'isica - Universidade Federal do Par\'a, Rua Augusto Corr\^ea n. 01, CEP 66075-110, Bel\'em, PA, Brazil}
\affil[2]{N\'ucleo Cosmo-Ufes - Universidade Federal do Esp\'irito Santo, A. Fernando Ferrari, n. 514, CEP 29075-910, Vit\'oria, ES, Brazil}
\affil[3]{Instituto de Engenharia e Geoci\^encias - Universidade Federal do Oeste do Par\'a, Rua Vera Paz, s/n, Bairro Sal\'e, CEP 68035-110, Santar\'em, PA, Brasil.}
\date{}

\begin{document}
	
\maketitle

\begin{abstract}
In a recent paper we studied the cosmology of Nonminimal Derivative Coupling (NDC) between gravity and a scalar field, which is a non-trivial class of Horndeski. We have shown that it presents a variety of solutions for the scale factor, but there are gravitational waves only for a very restrictive range in phase space, and primordial waves are completely forbidden classically in NDC. In this paper, we apply canonical quantization with the Bohm-de Broglie interpretation to that theory for a small value of the nonminimal coupling constant. We then study two quantum solutions of the Wheeler-DeWitt equation that lead to the perturbation of bouncing and cyclic solutions. By the analysis of the phase space obtained from the guidance equations of Bohm-de Broglie, we study the phase space determined by the scale factor and the scalar field for those quantum solutions in order to investigate three main aspects of that theory: the existence of non-singular solutions, the regions of accelerated expansion, and the regions compatible with the gravitational waves speed constraint. We conclude that there are non-singular solutions with an accelerated expansion period compatible with the constraint.\\

\end{abstract}


\section{Introduction}

The search of a working and coherent cosmological model that describes all stages of universe evolution has been the goal of virtually all theoretical cosmologists, for the last century. This objective deals with some serious challenges, specially for the very early universe.

As it is well known, inflationary theory was proposed in order to provide initial conditions for structure formation, preceding the radiation-dominated era \cite{pdg_review_2022}. Despite its enormous success, some authors argue that inflation can lead to some issues, like a fine-tuning potential \cite{turok_2014} and an initial singularity \cite{borde_e_vilenkin_1996,universe7120491}. To face such problems, there are a lot of proposals, like alternative theories of gravity and quantum cosmology theories.

In a previous paper \cite{ndc_class}, we reviewed one of the several aforementioned alternative theories: the so called Nonminimal Derivative Coupling (hereafter just called NDC), represented by the covariant Lagrangian density bellow:
\begin{equation}\label{lagr}
	L[g_{\mu\nu};\phi]=\sqrt{-g}\; \bigg[\frac{R}{8\pi}-g^{\mu\nu}\partial_{\mu}\phi\partial_{\nu}\phi-\kappa G^{\mu\nu}\partial_{\mu}\phi\partial_{\nu}\phi \bigg]\; ,
\end{equation}
where $R$ is the usual Ricci scalar of general relativity, $g_{\mu\nu}$ is the metric in a four dimensional spacetime, $G^{\mu\nu}$ is the Einstein tensor, and $\phi$ is a homogeneous scalar field. We consider flat Friedmann-Lema\^itre-Robertson-Walker (FLRW) metric
\begin{equation}\label{flrw}
	ds^2 =-N^2dt^2+ a^2\delta_{ij}dx^idx^j\; ,
\end{equation}
where $N(t)$ is the lapse function, which is a degree of freedom that comes from the spacetime foliation, that will become a Lagrange multiplier, thus providing a constraint which quantization gives the Wheeler-DeWitt equation.

The study of NDC and similar theories was motivated, among other reasons, by the fact that it is not a conformal equivalent to a canonical scalar field theory. In cosmology, it has been shown that NDC provides an asymptotic accelerated expansion in the very early universe, so it would be a candidate for inflation, without a fine tuning scalar potential \cite{PhysRevD.80.103505}. However, NDC has to deal with some other issues like the incompatibility with a canonical description of gravitational waves.

We reviewed NDC cosmology in a previous paper \cite{ndc_class}, which we briefly summarize in Section \ref{section_ndc_hamiltonian}, in order to point out the main properties of that theory. We have shown that it is very unlikely that NDC can describe gravitational waves at all, which would be trivial for general relativity with a minimally coupled scalar field, in contrast. This incompatibility has already been mentioned in a broader way by some previous reviews, like \cite{Kobayashi_2019,HEISENBERG20191}. Another feature of classical NDC is that its accelerated solution actually is asymptotically singular, thus it does not avoid a singularity, in principle.

But it is still possible that a quantum version of NDC could at the same time provide an accelerated expansion mechanism, avoid the Big Bang singularity (which could be a consequence of quantization), and also be compatible with the propagation of gravitational waves. Our goal in this paper is to construct a quantum version of NDC and investigate in what sense and to what extent that is possible, for two quantum solutions of the Wheeler-DeWitt equation of NDC.

In Section \ref{section_ndc_hamiltonian}, we point out that any attempt to construct a general hamiltonian version of \eqref{lagr} leads to a quintic algebraic equation, known to not possess a general solution in terms of elementary algebraic operations. Thus, we propose to perturbate over the coupling parameter $\kappa$, limiting all the analysis for the case $0<\kappa\ll 1$, which allows us to apply canonical quantization to the theory. This is compatible to \cite{PhysRevD.80.103505}, where $\kappa$ was constrained by the duration of inflation, leading to a very small value, despite the fact that the calculations therein where not linearized in $\kappa$. In other words, we are not getting too far from what is expected from NDC, in some sense.

Then, in Section \ref{section_quantizacao}, we apply canonical quantization to the constraint $\mathcal{H}\approx 0$, leading to the Wheeler-DeWitt equation of the system $\hat{\mathcal{H}}\psi = 0$, which we interpret in terms of Bohm-de Broglie quantum cosmology, following the theory presented in \cite{nelson_2005} and references therein. The quantum interpretation of quantum cosmology is an application of the quantum mechanics created by David Bohm and Louis de Broglie and its foundations can be found in the textbook \cite{HOLLAND199395}.

In Bohm-de Broglie, or Bohmian description, the wave function $\psi$ is a pilot-wave that guides the solution for the physical quantities of interest, so that a particular solution for the scale factor is given in a deterministic way by a dynamical system given by the so called guidance equations, as we shall see. The main reasons to apply a Bohmian interpretation is that it avoids the measurement conceptual issue and the problem of time, as mentioned in \cite{nelson_2005}.

Next, we present two pilot waves that solve the Wheeler-DeWitt equation, one that leads to bounce in Section \ref{section_bounce} and one that leads to cyclic solutions, in Section \ref{section_cyclic}. Both are obtained in a similar way: the original Wheeler-DeWitt is split into two equations perturbatively in $\kappa$, so the equation of order $\kappa^0$ has some well known bouncing and cyclic solutions, then we solve the $\kappa^1$ equations that perturbate the original bounces and cycles and present them as dynamical systems. For more about bouncing and cyclic solutions in a broader context, see, for instance, \cite{novello_2008_bouncing,steinhardt_prd_2002}.

In Section \ref{section_analysis}, we analyse those bouncing and cyclic solutions for the scale factor in order to answer the questions raised above, about the compatibility of NDC with gravitational waves, and if it can provide an accelerated expansion epoch without leading to a singularity. Finally, in Section \ref{section_conc} we make some conclusion remarks.

\section{Nonminimal Derivative Coupling Cosmology}\label{section_ndc_hamiltonian}

As shown in \cite{PhysRevD.80.103505} and reviewed in \cite{ndc_class}, the cosmology predicted by NDC presents an accelerated expansion phase $a(t) \sim e^{t/\sqrt{9\kappa}}$ when $\dot{a}/a>0$ and $|\dot{\varphi}|\rightarrow\infty$, which is the most important result, that could be a basis for an inflationary model. Other asymptotic limits of NDC are: a matter-domination when $\dot{\varphi}\rightarrow 0$; a stiff matter domination when $\kappa\rightarrow 0$; an accelerated contraction $a(t) \sim e^{-t/\sqrt{9\kappa}}$ when $\dot{a}/a<0$ and $|\dot{\varphi}|\rightarrow\infty$. Notice that all these classical solutions lead to or come from the singularity $a = 0$.

An issue with NDC and similar theories is their incompatibility with gravitational waves, whose squared speed $c_{gw}^{2}$ is determined by the Horndeski coefficients as:
\begin{equation}\label{c2gw}
	c_{\text{GW}}^{2}=\frac{G_4-X(\ddot{\phi}G_{5,X}+G_{5,\phi})}{G_4-2XG_{4,X}-X(H\dot{\phi}G_{5,X}-G_{5,\phi})} \; .
\end{equation}
In Horndeski theory, denoting the kinetic term by $X\equiv-\frac{1}{2}\nabla^{\mu}\phi\nabla_{\mu}\phi$, for $K(\phi,X)=2X$, $G_3=0$, $G_4=1/8\pi$ and taking any $G_5(\phi)$ such that $dG_5(\phi)/d\phi = \kappa$, we obtain precisely \eqref{lagr}.

For quantization purposes, we must write the Lagrangian in minisuperspace \cite{bojowald2011quantum}:
\begin{equation}\label{lagr_mini_alpha}
	L(\varphi,\alpha)=\frac{3e^{3\alpha}}{4\pi}\bigg(-\frac{\dot{\alpha}^2}{N} +\frac{\dot{\varphi}^2}{N}-\frac{3\kappa \dot{\alpha}^2\dot{\varphi}^2}{N^3}\bigg)\; ,
\end{equation}
where we have defined $\alpha\equiv\ln a$ and $\varphi\equiv\sqrt{4\pi/3}\;\phi$ which simplifies all the calculations. In \cite{ndc_class}, we studied $c_{gw}$ in NDC theory, and its value is given by, in terms of $\kappa$ and $\dot{\varphi}$:
\begin{equation}\label{cgw_geral}
	c_{gw}=\left(\frac{1-3\kappa\dot{\varphi}^2}{1+3\kappa\dot{\varphi}^2}\right)^{1/2} \; .
\end{equation}
From that, we deduced that there is a very restricted region in phase space for which $c_{gw}$ is real ($\sqrt{3\kappa}|\dot{\varphi}| < 1$), and an even smaller region ($\sqrt{3\kappa}|\dot{\varphi}| < \sqrt{30}\times 10^{-8}$) for which $c_{gw}$ is as near to 1 as the observations suggest.

Since Bohm-de Broglie formalism can be seen as an effective theory, because it is based on a modified Hamilton-Jacobi equation (with an effective potential), \eqref{cgw_geral} is still valid for the quantum version of NDC we will present here. The standard experimental constraint over the gravitational waves speed is given by \cite{PhysRevLett.119.161101,Goldstein_2017,Abbott_2017}:
\begin{equation}\label{constraint_ligo}
	1-3\times10^{-15}<c_{gw}\leq1+7\times10^{-16}\;.
\end{equation}

In \cite{ndc_class}, we have presented a comprehensive examination of the implications of \eqref{constraint_ligo} to NDC cosmology. In summary, we have shown NDC is not able to properly describe any accelerated solution with a real-valued $c_{gw}$ (see section 3 of \cite{ndc_class}). Thus, our conclusion was that the classical version of this theory basically cannot describe gravitational waves.

In order to study the derivative coupling from the quantum perspective we want to present, we have to begin by determining its Hamiltonian form. The canonical momenta associated with \eqref{lagr_mini_alpha} are:
	\begin{subequations}\label{mom_geral}
		\begin{align}
			p_{\alpha}&=-\frac{3e^{3\alpha}}{2\pi N}\dot{\alpha}\bigg(1+3\kappa\frac{\dot{\varphi}^2}{N^2}\bigg)\; ,\\
			p_{\varphi}&=\frac{3e^{3\alpha}}{2\pi N} \dot{\varphi}\bigg(1-3\kappa\frac{\dot{\alpha}^2}{N^2}\bigg)\; ,\\
			p_{N}&=0\; .
		\end{align}
	\end{subequations}
But a fundamental problem immediately arises: for a general $\kappa$, the system \eqref{mom_geral} leads to two quintic algebraic equations (or something even more complicated, depending on the method) when one tries to determine the generalized velocities as functions of the momenta. Thus, the Hamiltonian cannot be expressed in a closed analytical form, in terms of elementary operations, for the general case, as we know from Abel-Ruffini theorem \cite{lang2005algebra}. 
	
That is a really strong mathematical limitation, which makes us think that there is probably nothing left to be done with this idea. However, the inflationary model purposed in \cite{PhysRevD.80.103505} (which is the reason why we consider NDC theory in cosmology) requires a small value of $\kappa$. Also, according to \cite{ndc_class}, the gravitational waves speed constraint indicates that the smaller the value of $\kappa$ is, the wider the range of allowed values of $\dot{\varphi}$ becomes.


Hence, we can consider a small value of the nonminimal coupling constant $\kappa$. Since NDC describes an inflationary model only for a positive $\kappa$ (as shown in \cite{PhysRevD.80.103505}), we can then take $0<\kappa\ll 1$. This allows us to see the nonminimal coupling term as a small contribution to the others, with $\kappa$ being the perturbative parameter.
	
	
When $0<\kappa\ll 1$, system \eqref{mom_geral} can be solved for the generalized velocities as functions of the momenta:
	\begin{subequations}\label{veloc_gen_kpeq}
		\begin{align}
			\dot{\alpha}&=-\frac{2\pi}{3}e^{-3\alpha}N p_{\alpha}\left(1-\frac{4\pi^2\kappa}{3}e^{-6\alpha}p_{\varphi}^{2}\right)\; ,\\
			\dot{\varphi}&= \frac{2\pi}{3}e^{-3\alpha}N p_{\varphi}\left(1+\frac{4\pi^2\kappa}{3} e^{-6\alpha}p_{\alpha}^{2}\right)\; .
		\end{align}
	\end{subequations}
Therefore, applying the usual Legendre transformation $H=\sum \dot{q}_jp_j-L$, the Hamiltonian $H$ linearized in $\kappa$ is given by:
	\begin{equation}\label{ham}
		H=\frac{N\pi e^{-3\alpha}}{3}\left( -p_{\alpha}^{2}+p_{\varphi}^{2}+\frac{4\pi^2\kappa}{3}\pi^2 e^{-6\alpha} p_{\alpha}^{2}p_{\varphi}^{2} \right) \equiv N\mathcal{H}\;,
	\end{equation}
where $\mathcal{H}$ does not depend on $N$. It thus follows that $H$ is a constrained Hamiltonian, so that $N$ plays the role of a Lagrange multiplier.

\section{Canonical Quantization}\label{section_quantizacao}
	
We can now apply Dirac quantization rule to the perturbed Hamiltonian above, which leads to the following Wheeler-DeWitt (WDW) equation, $\hat{\mathcal{H}}\psi=0$:
	\begin{equation}\label{eqwdw_ini}
		\frac{\partial^2\psi}{\partial\alpha^2}-\frac{\partial^2\psi}{\partial\varphi^2}=-\frac{4\pi^2\hbar^2\kappa}{3}e^{-6\alpha}\frac{\partial^4\psi}{\partial\alpha^2\partial\varphi^2}\; .
	\end{equation}
As mentioned in the Introduction, we are going to solve \eqref{eqwdw_ini} and interpret its solutions by means of the Bohm-de Broglie quantization. The usual procedure is to write $\psi=R\, e^{iS/\hbar}$, which leads to two equations, which correspond to the real and imaginary parts. The real part is analogous to a Hamilton-Jacobi equation from classical mechanics, from which we postulate the guidance equations and we also define the quantum potential. Let us then apply this method, but also perturbating all equations to be linearized in $\kappa$.

Now, denoting the wave function up to first order in $\kappa$ as
	\begin{equation}\label{psi_linearized}
		\psi(\varphi,\alpha)=\psi_0(\varphi,\alpha)+\kappa\,\psi_1(\varphi,\alpha) \; ,
	\end{equation}
we can split \eqref{eqwdw_ini} into two equations:
	\begin{subequations}\label{eqwdw_sepk}
		\begin{align}
			\frac{\partial^2\psi_0}{\partial\alpha^2}-\frac{\partial^2\psi_0}{\partial\varphi ^2} &=0\; ,\label{wdwo0}\\
			\frac{\partial^2\psi_1}{\partial\alpha^2}-\frac{\partial^2\psi_1}{\partial\varphi ^2} &=-\frac{4\pi^2\hbar^2}{3}e^{-6\alpha}\frac{\partial^4\psi_0}{\partial\alpha^2\partial\varphi ^2}\; .\label{wdwo1}
		\end{align}
	\end{subequations}
In the two next sections, we will solve these equations in the following way: we take a known quantum-cosmological solution to \eqref{wdwo0}, which will provide the second part of \eqref{wdwo1}, thus making it a non-homogeneous equation. With both solutions, we have a complete wave function according to \eqref{psi_linearized}, as it is usual for such perturbation methods.

Next, writing $\psi=R\, e^{iS/\hbar}$, where $R$ and $S$ are real and 
	\begin{subequations}\label{rsrsrs}
		\begin{align}
			R&=R_0+\kappa R_1\; ,\\
			S&=S_0+\kappa S_1\; ,
		\end{align}
	\end{subequations}
we can obtain $\psi_0$ and $\psi_1$ in terms of $R$ and $S$ terms:
	\begin{subequations}\label{psi-expans-rs-gen}
		\begin{align}
			\psi_0&=R_0e^{iS_0/\hbar}\; ,\\
			\psi_1&=\bigg(R_1+\frac{i}{\hbar}R_0S_1\bigg)e^{iS_0/\hbar}\; .
		\end{align}
	\end{subequations}	
	
Then, inserting $\psi=R\, e^{iS/\hbar}$ into \eqref{eqwdw_ini}, its real part can be written as
	\begin{align}\label{hj_unpert}
		&e^{-3\alpha}\bigg[ -\frac{\pi}{3}\bigg(\frac{\partial S}{\partial\alpha}\bigg)^2 + \frac{\pi}{3}\bigg(\frac{\partial S}{\partial\varphi}\bigg)^2+\frac{4\pi^3}{9}\kappa e^{-6\alpha} \bigg(\frac{\partial S}{\partial\alpha}\frac{\partial S}{\partial\varphi}\bigg)^2\bigg]+Q(\varphi,\alpha)=0\; ,
	\end{align}
or, more precisely, expanding in $\kappa$, 
	\begin{subequations}
		\begin{align}
			& e^{-3\alpha}\bigg[ -\frac{\pi}{3}\bigg(\frac{\partial S_0}{\partial\alpha}\bigg)^2 + \frac{\pi}{3}\bigg(\frac{\partial S_0}{\partial\varphi}\bigg)^2\bigg]+Q_0(\varphi,\alpha)=0\; ,\label{hj_pert0}\\
			& e^{-3\alpha}\bigg[-\frac{2\pi}{3}\frac{\partial S_0}{\partial\alpha}\frac{\partial S_1}{\partial\alpha}+ \frac{2\pi}{3}\frac{\partial S_0}{\partial\varphi}\frac{\partial S_1}{\partial\varphi}
			+\frac{4\pi^3}{9}e^{-6\alpha}\bigg(\frac{\partial S_0}{\partial\alpha} \frac{\partial S_0}{\partial\varphi}\bigg)^2  \bigg]+\nonumber\\ 
			&+Q_1(\varphi,\alpha)=0\; ,\label{hj_pert1}
		\end{align}
	\end{subequations}
where $Q=Q_0+\kappa Q_1$, with
	\begin{equation}
		Q_0=\frac{\pi\hbar^2e^{-3\alpha}}{3R_0}\bigg(\frac{\partial^2R_0}{\partial\alpha^2} - \frac{\partial^2R_0}{\partial\varphi^2} \bigg)
	\end{equation}
and
	\begin{align}
		Q_1 &=\frac{\pi\hbar^2e^{-3\alpha}}{3R_0}\bigg[\frac{R_1}{R_0}\bigg(\frac{\partial^2R_0}{\partial\varphi^2}-\frac{\partial^2R_0}{\partial\alpha^2}\bigg) +\frac{\partial^2R_1}{\partial\alpha^2} - \frac{\partial^2R_1}{\partial\varphi^2}\bigg]\nonumber\\
		&-\frac{4\pi^3\hbar^2e^{-9\alpha}}{9R_0}\bigg[2\frac{\partial^2S_0}{\partial\alpha^2}\frac{\partial S_0}{\partial\varphi}\frac{\partial R_0}{\partial\varphi}+2\frac{\partial^2S_0}{\partial\varphi^2}\frac{\partial S_0}{\partial\alpha}\frac{\partial R_0}{\partial\alpha}\nonumber\\
		&+\frac{\partial^2R_0}{\partial\alpha^2}\bigg(\frac{\partial S_0}{\partial\varphi}\bigg)^2+\frac{\partial^2R_0}{\partial\varphi^2}\bigg(\frac{\partial S_0}{\partial\alpha}\bigg)^2\nonumber\\
		&+4\frac{\partial^2S_0}{\partial\alpha\partial\varphi}\bigg(\frac{\partial R_0}{\partial\alpha} \frac{\partial S_0}{\partial\varphi} + \frac{\partial R_0}{\partial\varphi} \frac{\partial S_0}{\partial\alpha}\bigg) + 4\frac{\partial S_0}{\partial\alpha}\frac{\partial S_0}{\partial\varphi} \frac{\partial^2R_0}{\partial\alpha\partial\varphi}\bigg] \nonumber\\
		&-\frac{8\pi^3\hbar^2e^{-9\alpha}}{9}\bigg[\frac{\partial S_0}{\partial\alpha} \frac{\partial^3S_0}{\partial\alpha\partial\varphi^2} + \frac{\partial S_0}{\partial\varphi} \frac{\partial^3S_0}{\partial\alpha^2\partial\varphi}+\bigg( \frac{\partial^2S_0}{\partial\alpha\partial\varphi} \bigg)^2 \bigg] \nonumber\\
		&-\frac{4\pi^3\hbar^2e^{-9\alpha}}{9}\frac{\partial^2S_0}{\partial\alpha^2}\frac{\partial^2S_0}{\partial\varphi^2} + \frac{4\pi^3\hbar^4e^{-9\alpha}}{9R_0}\frac{\partial^4R_0}{\partial\alpha^2\partial\varphi^2}\; .\label{q1_geral}
	\end{align}
	
Eq. \eqref{hj_unpert} has a form analogous to a Hamilton-Jacobi equation, with $S$ playing the role of the Hamilton principal function associated with $H$ and interpreting $Q$ as a potential, since it does not depend on kinetic terms, as we shall explain bellow.

Now we can construct a Bohm-de Broglie theory in two steps. The first step is to generalize the classical relation between $S$ and the momenta, which are
	\begin{equation}\label{eq_guias_mom}
		p_{\alpha}=\frac{\partial S}{\partial\alpha}\qquad\mbox{and}\qquad p_{\varphi}=\frac{\partial S}{\partial\varphi}\; .
	\end{equation}
These are called the \textit{guidance equations}. Here lies the solution for the problem of time in Bohmian theory: those equations are invariant under a redefinition of time \cite{acacio_nelson_problema_do_tempo_e_sing}. Besides that, in practice the absence of time in \eqref{eqwdw_ini} is not problematic in this interpretation because inserting guidance equations \eqref{eq_guias_mom} into \eqref{veloc_gen_kpeq} gives (setting now $N=1$):
	\begin{subequations}\label{eq_guias_vel}
		\begin{align}
			\dot{\alpha} &=-\frac{2\pi }{3}e^{-3\alpha}\Bigg\{\frac{\partial S_0}{\partial\alpha}+\kappa\Bigg[\frac{\partial S_1}{\partial\alpha}-\frac{4\pi^2}{3} e^{-6\alpha}\frac{\partial S_0}{\partial\alpha}\Bigg(\frac{\partial S_0}{\partial\varphi}\Bigg)^2\Bigg]\Bigg\}	\; ,		\label{eq_guia_alpha}\\
			\dot{\varphi} &=\frac{2\pi }{3}e^{-3\alpha}\Bigg\{\frac{\partial S_0}{\partial\varphi}+\kappa\Bigg[\frac{\partial S_1}{\partial\varphi}+\frac{4\pi^2}{3} e^{-6\alpha}\frac{\partial S_0}{\partial\varphi}\Bigg(\frac{\partial S_0}{\partial\alpha}\Bigg)^2\Bigg]\Bigg\}	\; ,		\label{eq_guia_phi}
		\end{align}
	\end{subequations}
which, for a given solution $\psi$, becomes a dynamical system that allows us to study the evolution of $\alpha(t)$ and $\varphi(t)$, for given initial conditions. It is worth remembering that Bohm-de Broglie theory is deterministic and this fact becomes evident thanks to the new form of the guidance equations \eqref{eq_guias_vel}.
	
The quantity $Q$ in \eqref{hj_unpert} does not depend on the derivatives of $S$ perturbatively, as we can see from \eqref{q1_geral}, in the sense that $Q_0$ does not depend on $\partial_iS_0$ (nor $\partial_iS_1$) and $Q_1$ does not depend on $\partial_iS_1$. Based on this fact and also on Eq. \eqref{eq_guias_mom}, we can say that $Q$ is not a true kinetic term. The second step to implement de Broglie-Bohm, which is tightly connected with the first one, is then to recognize $Q$ as the \textit{quantum potential}, which in the present case also corresponds to the effective potential.
	
The concept of quantum potential is of fundamental relevance in this description, because the quantum version of Hamilton-Jacobi Eq. \eqref{hj_unpert} shows us that $Q=0$ implies that the dynamics is purely classical. In other words, $Q\neq0$ means a quantum contribution. Hence, $Q(\varphi,\alpha)$ maps quantum effects over the phase space. In particular, the classical correspondence is given by the limit $Q\rightarrow0$, for a particular solution $\psi$.
	
In summary, to actually apply this algorithm, we will solve \eqref{eqwdw_sepk}, thus obtaining the wave function \eqref{psi_linearized}, from which we obtain a dynamical system \eqref{eq_guias_vel} to be studied in the phase space $\varphi\times\alpha$. This last step is easier if we notice that
\begin{equation}\label{eq_guias_mom2}
	\frac{\partial S}{\partial\alpha}=\hbar\;{\rm Im} \; \left(\frac{\partial_{\alpha}\psi}{\psi}\right) \qquad\mbox{and}\qquad \frac{\partial S}{\partial\varphi}=\hbar\;{\rm Im} \; \left(\frac{\partial_{\varphi}\psi}{\psi}\right) \; ,
\end{equation}
which follows from the differentiation of $\psi=R\, e^{iS/\hbar}$.



\section{Numerical approximation of critical points}

In the next Sections, we will study the dynamical system \eqref{eq_guias_vel} for two pilot-waves, which will provide us with the description of quantum cosmic evolution. In this sense, knowing the type and exact position of the system's critical points is essential. With this in mind, and considering the 
complexity of obtaining an analytical solution for these points from the solution of (12) (which we will comment in the next Section), we define a functional that has the same critical points as the original dynamical system and is numerically easier to approximate. The chosen functional is
\begin{equation}
	T(\alpha,\varphi) = \dot{\alpha}^2(\alpha,\varphi) + \dot{\varphi}^2(\alpha,\varphi) \; ,
	\label{func_root}
\end{equation}
which is identically zero when $ \dot{\alpha} = \dot{\varphi} = 0 $. The numerical approximation of the critical points of \eqref{func_root} can be easily found through an iterative numerical minimization process, which generally uses the direction of the gradient vector of the functional to be minimized \cite{nocedal2006numerical}. 
In Mathematica\textregistered{} software \cite{wolfram2019mathematica}, there are specialized functions for this type of search, such as \texttt{FindArgMin(args)} or 
\texttt{FindArgMax(args)},  which given the function \texttt{f(x,y)} and an initial point ($ x_o,y_o $) near a local or global maximum or minimum, return the coordinates of these critical points with excellent precision. 

\begin{figure}[h]  
	\centering      
	\includegraphics[scale=0.65]{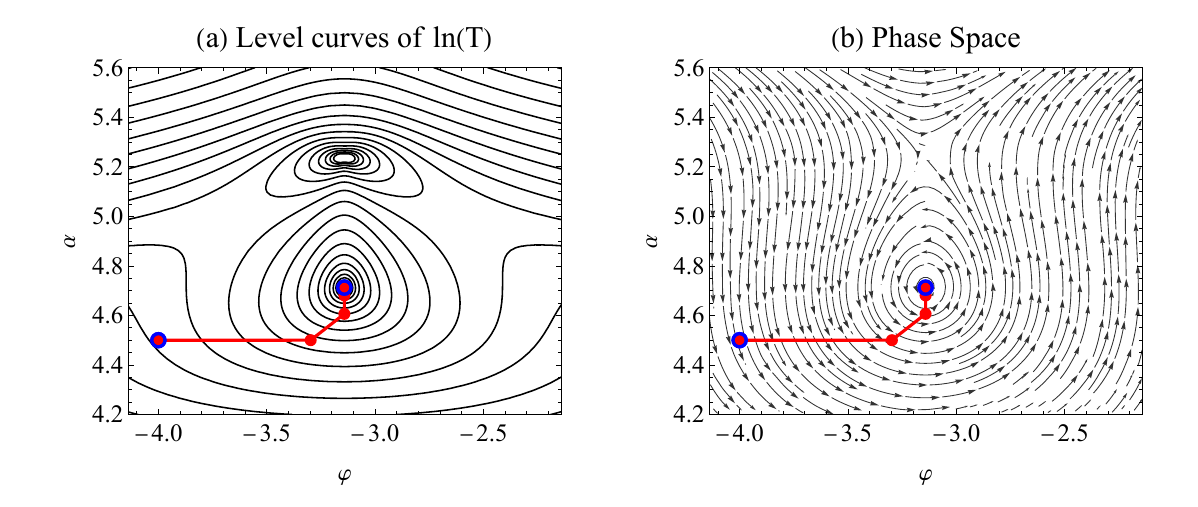}
	\caption{Representation of (a) the functional to be minimized and (b) the phase space near the critical points of the dynamic system.}   
	\label{fig1}       
\end{figure}

To exemplify the use of the \texttt{FindArgMin(args)} function, Figure \ref{fig1} shows the convergence to a critical point from Figure \ref{fig3},
for $\kappa = 10^{-2}$, near the coordinates $\alpha = 4.7$ and $\varphi = -3.14$, starting the iterative process from the point 
$\alpha_o = 4.5$ and $\varphi_o = -4.5$. The blue circular points indicate the initial and final points of the iterative process and the
solid red points represent the intermediate points of the method's convergence.

The critical points obtained in the phase spaces shown in Figures \ref{fig2} and \ref{fig3} were determined using these functions, according to the specifics of each case.

\section{Perturbed Bounce Solutions}\label{section_bounce}
Let us first set units such that $\hbar=1$. Given any solution $\psi_0$ of \eqref{wdwo0}, Eq. \eqref{wdwo1} becomes a non-homogeneous second order linear PDE, and this means that the actual solution for $\psi$ has the form
\begin{equation}
	\psi=\psi_0+\kappa(\psi_{\rm 1h}+\psi_{\rm 1p})	\; ,	\label{psi_varphi_10}
\end{equation}
where $\psi_{\rm 1h}$ and $\psi_{\rm 1p}$ are, respectively, the homogeneous and the particular solutions for Eq. \eqref{wdwo1}. But $\psi_{\rm 1h}$ is just $\psi_{\rm 0}$, so that if some  $\psi_0$ is given, we just have to determine the particular solution $\psi_{\rm 1p}$. This means that we can study a variety of solutions, one for each given $\psi_{\rm 0}$.

In this Section, we consider as $\psi_0$ the bounce solution presented in \cite{PhysRevD.57.4707,PINTONETO2000194,Colistete:2000ix} and reviewed in \cite{PhysRevD.97.083517}:
\begin{align}
	\psi_0 &= \sqrt{\pi} \left[e^{-\frac{1}{4} (\alpha
		-\varphi ) (\alpha -\varphi -4 i)}+e^{-\frac{1}{4}
		(\alpha +\varphi ) (\alpha +\varphi +4
		i)}\right]\; ,\label{psi0nelson}
\end{align}
which is the result of integrating a Gaussian wave packet.
	
Eq. \eqref{wdwo1} has no easy direct solution when $\psi_0$ is given by \eqref{psi0nelson}, but we can split \eqref{wdwo1} into two simpler equations, as follows. First, let us introduce the variables

\begin{equation}\label{mud_var_xy}
	x\equiv \varphi-\alpha \qquad \mbox{and} \qquad y \equiv \varphi+\alpha \; ,
\end{equation}
which allows us to rewrite \eqref{wdwo1}, after some manipulation, as:
\begin{equation}\label{wdw01_xy}
	\partial_x\partial_y\psi_{1p}=\textstyle\frac{\pi^2}{3}e^{3x-3y}\hat{L}\psi_0 \;.
\end{equation}
where $\hat{L}\equiv\partial^4_x-2\partial^2_x\partial^2_y+\partial^4_y$ is a linear differential operator. The advantage of this substitution is that now the inhomogeneities can be written as a sum such that the resulting equations are separable.

In terms of $x$ and $y$, the 0th order wave function \eqref{psi0nelson} can be written as:
\begin{equation}
	\psi_0(x,y)=\psi_{0x}(x)+\psi_{0y}(y)\equiv \sqrt{\pi}e^{-ix-x^2/4}+\sqrt{\pi}e^{-iy-y^2/4} \; .
\end{equation}
Thus, because of a well known property of non-homogeneous equations, it suffices to consider $\psi_{1p}$ has the form:
\begin{equation}\label{psi1pfxfy}
	\psi_{1p}(x,y) = f_{1}(x,y) + f_{2}(x,y) \; , 
\end{equation}
where $f_{1}$ and $f_{2}$ are defined as the solutions of the equations below:
\begin{subequations}\label{sistema_fxy}
	\begin{align}
		\partial_x\partial_y f_{1}(x,y) &=\textstyle\frac{\pi^2}{3}e^{3x-3y}\partial^4_x\psi_{0x}(x) \; ,\\
		\partial_x\partial_y f_{2}(x,y) &=\textstyle\frac{\pi^2}{3}e^{3x-3y}\partial^4_y\psi_{0y}(y)\; .
	\end{align}
\end{subequations}

If we now consider that the unknown functions $f_1$, $f_2$ can be written as
\begin{subequations}\label{fxfybounce}
	\begin{align}
		f_{1}(x,y) &= \frac{\pi^2}{3}e^{-3y}F_{1}(x) \; ,\\
		f_{2}(x,y) &= \frac{\pi^2}{3}e^{3x}F_{2}(y) \; ,
	\end{align}
\end{subequations}
we are lead to
\begin{subequations}\label{FxFy_bounce}
	\begin{align}
		F_{1}'(x) &= -\frac{\sqrt{\pi}}{48}e^{3x-ix-x^2/4}\left( 76-80ix-36x^2+8ix^3+x^4 \right) \; ,\\
		F_{2}'(y) &= \frac{\sqrt{\pi}}{48}e^{-3y-iy-y^2/4}\left( 76-80iy-36y^2+8iy^3+y^4 \right) \; ,
	\end{align}
\end{subequations}
whose solutions are, setting all integration constants to zero for simplicity:
\begin{subequations}\label{FxFy_bounce_sol}
	\begin{align}
		F_{1}(x) &=  \frac{\sqrt{\pi}}{24} \exp\left[(3-i)x-\frac{x^2}{4} \right] \left[ 180 + 52 i + 6(3+4i)x +6(1+i)x^2 +x^3 \right] \nonumber \\
				 &-27\pi e^{8-6i} \text{erf}\left( i-3+\frac{x}{2} \right) \; ,\\
		F_{2}(y) &=  \frac{\sqrt{\pi}}{24} \exp\left[ -(3+i)y-\frac{y^2}{4} \right] \left[ 180 - 52 i - 6(3-4i)y +6(1-i)y^2 -y^3 \right] \nonumber \\
				 &+ 27\pi e^{8+6i} \text{erf}\left( i+3+\frac{y}{2} \right) \; ,
	\end{align}
\end{subequations}
where $\text{erf}$ the usual error function defined by $\text{erf}(z)=\frac{2}{\sqrt{\pi}}\int_{0}^{z}e^{-t^2}dt$ for any $z\in\mathbb{C}$.

Now, inserting these solutions into \eqref{fxfybounce} and coming back to the variables $\alpha$ and $\varphi$, we obtain $\psi_{1p}$ as the sum \eqref{psi1pfxfy}, which leads to
\begin{align}
	\psi_{1p}(\varphi,\alpha)&=\frac{\pi ^{5/2}}{72} \kappa  e^{-3 (\alpha
		+\varphi )} \Big\{e^{-\frac{1}{4} (\alpha -\varphi )
		(\alpha -\varphi +12-4i)} [180+52i\nonumber\\
	&+(18+24i) (\varphi -\alpha )+6(1+i)(\varphi - \alpha
	)^2+(\varphi -\alpha
	)^3]\nonumber\\
	&-648 e^{8-6 i} \sqrt{\pi }\; 
	\text{erf}\Big(\frac{\varphi-\alpha}{2}-3+
	i\Big)\nonumber\\
	&-e^{6 \varphi -\frac{1}{4} (\alpha +\varphi )
		[\alpha +\varphi +12+4i]} \Big[52i-180 + (18-24i)(\alpha+\varphi)\nonumber\\
	&+6(i-1) (\alpha +\varphi )^2 + (\alpha +\varphi)^3\nonumber\\
	&-648 \sqrt{\pi } e^{\frac{1}{4}(\alpha+\varphi +6+2i)^2}\text{erf}\Big(\frac{\varphi+\alpha}{2} +3+i\Big)\Big]\Big\} \; .
\end{align}
Therefore, the complete wave function is
\begin{align}\label{psi_bounce}
	\psi(\varphi,\alpha)&= \sqrt{\pi } (1+\kappa) \left[e^{-\frac{1}{4} (\alpha
		-\varphi ) (\alpha -\varphi -4 i)}+e^{-\frac{1}{4}
		(\alpha +\varphi ) (\alpha +\varphi +4
		i)}\right] +\psi_{1p}(\varphi,\alpha) \; .
\end{align}

In order to actually study the quantum dynamics of \eqref{psi_bounce}, due to the presence of the error function ${\rm erf}(z)$ and the guidance equations, the following well-known identities are useful:
\begin{subequations}\label{identities_erf}
	\begin{align}
		{\rm erfi}(z) &= -i\, {\rm erf} (iz)\; , \label{erfi_def}\\
		{\rm Re}\left[ {\rm erf} (x+iy) \right] &= {\textstyle \frac{1}{2} }\left[ {\rm erf} (x+iy) + {\rm erf} (x-iy) \right] \; , \\
		{\rm Im}\left[ {\rm erf} (x+iy) \right] &= {\textstyle \frac{1}{2i} }\left[ {\rm erf} (x+iy) - {\rm erf} (x-iy) \right] \; ,
	\end{align}
\end{subequations}
for any $x,y\in\mathbb{R}$, $z\in\mathbb{C}$. Here, \eqref{erfi_def} is the definition of an auxiliary map called the imaginary error function, which can also be used to simplify the calculations.

\begin{figure}[h]
	\centering
	\includegraphics[width=\linewidth]{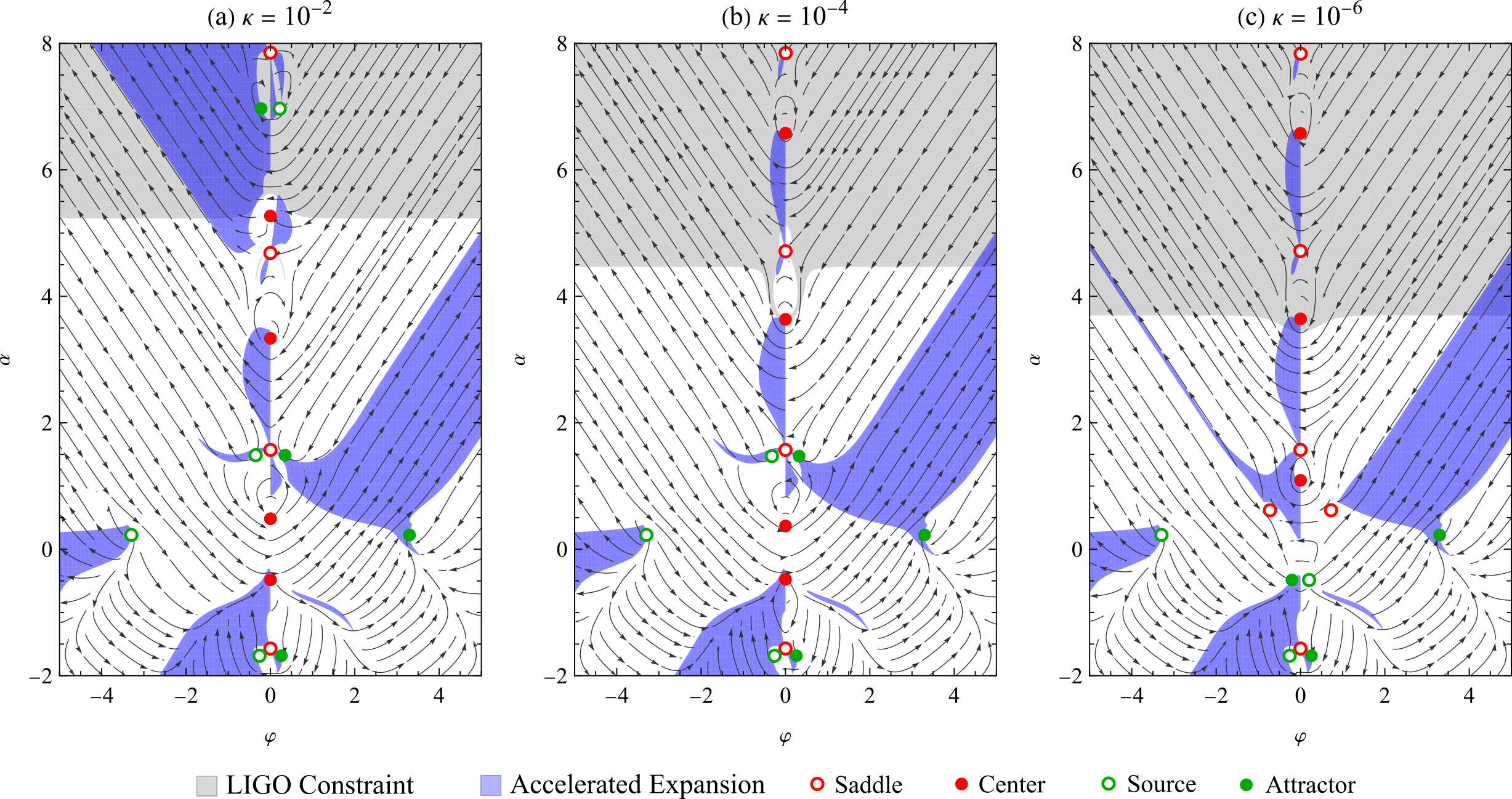}\caption{Phase space for the bounce solution \eqref{psi_bounce} for three decreasingly small values of the perturbation parameter $\kappa$. The gray shaded regions represent the points compatible with the gravitational wave constraint \eqref{constraint_ligo} and the blue regions represent accelerated expansion ($\ddot{a},\dot{a}>0$). The stream lines are quantum solutions of the guidance equations \eqref{eq_guias_vel}. The orientation of the stream lines represent time evolution. We also show all four types of critical points: attactors, sources, saddle points and centers.}
	\label{fig2}	
\end{figure}

Finally, inserting \eqref{psi_bounce} into \eqref{eq_guias_vel} up to the first order in $\kappa$, with the help of the identities \eqref{identities_erf}, one can obtain a dynamical system of the form
\begin{subequations}\label{din_sys}
	\begin{align}
		\dot{\alpha}_{\kappa}(\varphi,\alpha) &= f^{0}(\varphi,\alpha)+\kappa f^{1}(\varphi,\alpha) \; , \\
		\dot{\varphi}_{\kappa}(\varphi,\alpha)  &= g^{0}(\varphi,\alpha)+\kappa g^{1}(\varphi,\alpha) \; ,
	\end{align}
\end{subequations}
for some $f^{i},g^{i}$. The actual analytical expression of \eqref{din_sys} is very complicated, thanks to \eqref{eq_guias_mom2}: one must differentiate $\psi$ with respect to $\alpha$ and $\varphi$, divide the result by $\psi$, then take the imaginary part of the quotient and expand the result up to the first order in $\kappa$. This process leads to hundreds of terms for $f^{1}$ and $g^{1}$, so we provide the analytical form of \eqref{din_sys} as a separate file and analyse its cosmological implications in Section \ref{section_analysis}. The resulting phase portrait is shown in Figure \ref{fig2}.


\section{Perturbed Cyclic Solutions}\label{section_cyclic}

In this Section, we will consider as the 0th order solution:
\begin{equation}\label{psi0_ciclica}
	\psi_0(\varphi,\alpha) = 1+2e^{i\omega\varphi}\cos (\omega\alpha) \; ,
\end{equation}
where $\omega$ is a real constant. This solution is adapted from a similar one presented in \cite{Torres_2020}, and it leads to cyclic universes, even for the $\kappa^{0}$ solution. We can solve the corresponding Wheeler-DeWitt equation \eqref{eqwdw_sepk} by a process similar to the one implemented in the previous Section.

We start from the 0th order solution \eqref{psi0_ciclica} and insert it into \eqref{wdwo1}, which leads to
\begin{equation}\label{wdw_o1_af_ciclica}
	\frac{\partial^2\psi_{1p}}{\partial\alpha^2}-\frac{\partial^2\psi_{1p}}{\partial\varphi ^2}= -\frac{8}{3}\pi^2\hbar^2\omega^4e^{-6\alpha+i\omega\varphi} \cos \omega\alpha \; .
\end{equation}
In terms of the variables $x$ and $y$ defined in \eqref{mud_var_xy}, this becomes:
\begin{equation}\label{wdw_ciclica_xy}
	-4\partial_x\partial_y\psi_{1p}=g_1(x,y)+g_2(x,y) \; ,
\end{equation}
where (again taking natural units such that $\hbar=1$):
\begin{subequations}
\begin{align}
	g_1(x,y) & = -\frac{8}{3}\pi^2\omega^4e^{3x+i\omega x/2}\cos\frac{\omega x}{2}e^{-3y+i\omega y/2}\cos\frac{\omega y}{2} \; ,\label{def_g1_ciclica} \\
	g_2(x,y) & = -\frac{8}{3}\pi^2\omega^4e^{3x+i\omega x/2}\sin\frac{\omega x}{2}e^{-3y+i\omega y/2}\sin\frac{\omega y}{2} \; .\label{def_g2_ciclica}
\end{align}
\end{subequations}

Note that again \eqref{wdw_o1_af_ciclica} is a non-homogeneous equation, so we can split it into two, just by writing $\psi_{1p}$ as a sum 
\begin{equation}\label{separacao_f12_ciclica}
	\psi_{1p}(x,y)=f_1(x,y)+f_2(x,y) \; ,
\end{equation}
where each $f_j$ satisfies 
\begin{equation}\label{wdw_ciclica_xy_separada}
	-4\partial_x\partial_y f_j = g_j \; ,
\end{equation}
since we can sum up \eqref{wdw_ciclica_xy_separada} in $j\in\{1,2\}$ and obtain \eqref{wdw_ciclica_xy}, as required.

Now, each $g_j$ can be written as 
\begin{equation}
	g_j(x,y) = -\frac{8}{3}\pi^2\omega^4 u_j(x) v_j(y) \; ,
\end{equation}
so that \eqref{wdw_ciclica_xy_separada} is separable. Separating each function $f_j$ as 
\begin{equation}
	f_j(x,y)=f_{jx}(x)f_{jy}(y) \; ,
\end{equation}
we obtain a total of four ODEs (with a separation constant $\lambda\neq 0$ that will cancel out later):
\begin{subequations}
	\begin{align}
		f_{1x}'(x) &= \lambda e^{3x+i\omega x/2}\cos\Big(\frac{\omega x}{2}\Big) \; , \\
		f_{1y}'(y) &=\frac{2\pi^2\omega^4}{3\lambda}e^{-3y+i\omega y/2}\cos\Big(\frac{\omega y}{2}\Big) \; , \\
		f_{2x}'(x) &=\lambda e^{3x+i\omega x/2}\sin\Big(\frac{\omega x}{2}\Big) \; , \\
		f_{2y}'(y) &= \frac{2\pi^2\omega^4}{3\lambda}e^{3x+i\omega x/2}\sin\Big(\frac{\omega x}{2}\Big) \; ,
	\end{align}
\end{subequations}
whose solutions are:
\begin{subequations}
\begin{align}
	f_{1x}(x) &=\frac{\lambda}{6}e^{3x}\left( 1+\frac{3e^{i\omega x}}{3+i\omega} \right) \; , \\
	f_{1y}(y) &=-\frac{1}{9\lambda}\pi^2\omega^4e^{-3y}\left( 1 + \frac{3e^{i\omega y}}{3 - i\omega} \right) \; , \\
	f_{2x}(x) &=\frac{i\lambda}{6}e^{3x}\left( 1 - \frac{3e^{i\omega x}}{3+i\omega} \right) \; , \\
	f_{2y}(y) &=-\frac{i}{9\lambda}\pi^2\omega^4e^{-3y}\left( 1 - \frac{3e^{i\omega y}}{3-i\omega} \right) \; .
\end{align}
\end{subequations}

Finally, from the relations above, we can write down the particular solution, after some simplifications, as:
\begin{equation}
	\psi_{1p}(\varphi,\alpha)=\frac{2\pi^2\omega^4}{9(9+\omega^2)}e^{-6\alpha+i\omega\varphi}\left[ \omega \sin \omega \alpha-3\cos \omega\alpha \right] \; .
\end{equation}
Therefore, in this case, the final solution $\psi$ is:
\begin{equation}\label{psi_ciclica}
	\psi(\varphi,\alpha)=(1+\kappa)\left(1+2e^{i\omega\varphi}\cos \omega\alpha \right) + \kappa\frac{2\pi^2\omega^4}{9(9+\omega^2)}e^{-6\alpha+i\omega\varphi}\left[ \omega \sin \omega \alpha-3\cos \omega\alpha \right] \; .
\end{equation}

\begin{figure}[h]
	\centering
	\includegraphics[width=\linewidth]{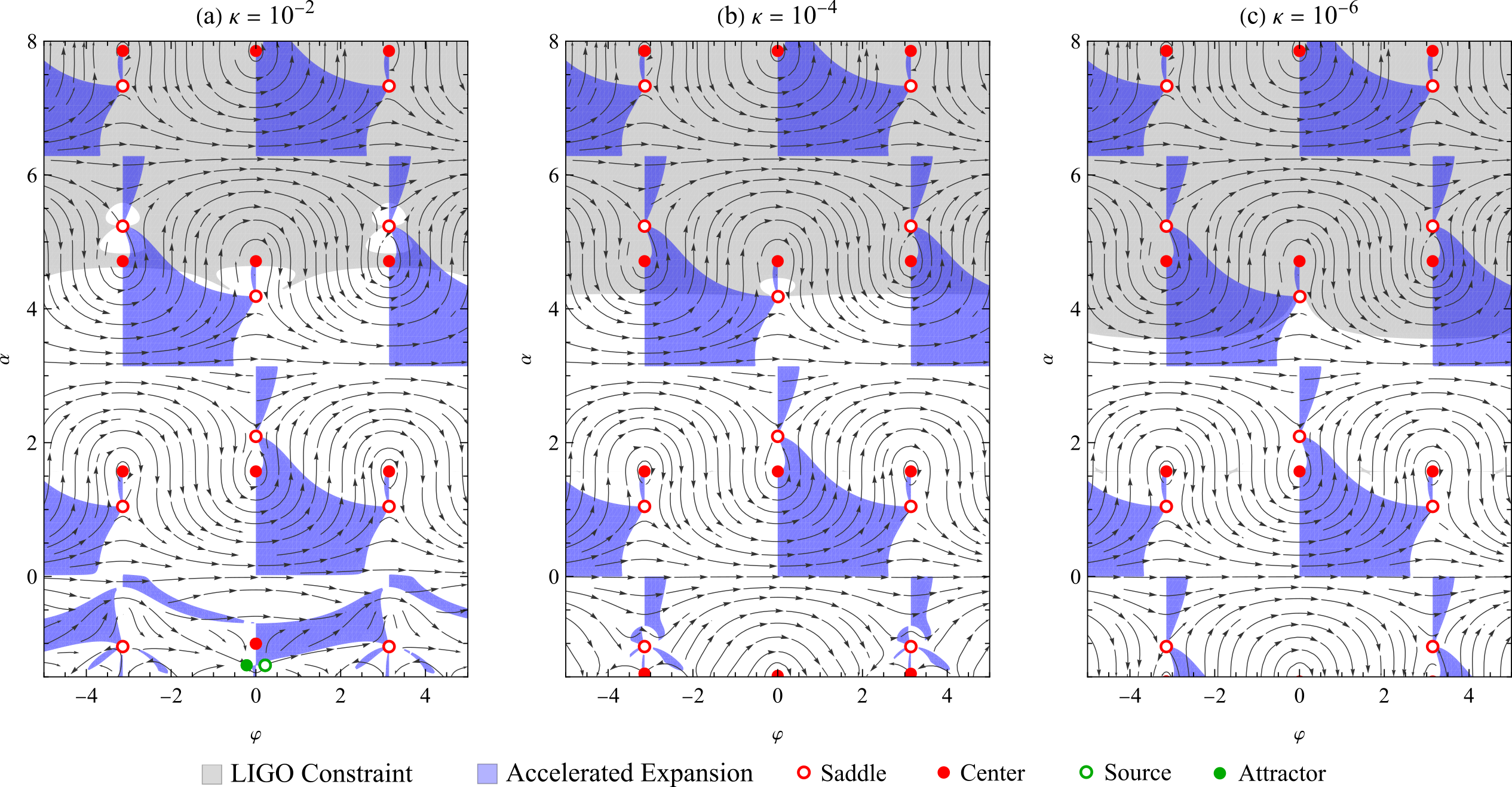}
	\caption{Phase portrait for the cyclic solution \eqref{psi_ciclica}, taking $\omega = 1$, for three decreasingly small values of $\kappa$. The shaded regions, stream lines and critical points all follow the same convention of Figure \ref{fig2}, but for the dynamical system obtained from the cyclic solution \eqref{psi_ciclica}.}
	\label{fig3}
\end{figure}

We will interpret this pilot-wave and the previous one in the next Section and the phase portrait obtained from it is shown in \ref{fig3}.


\section{Analysis of the Solutions}\label{section_analysis}

From the bouncing \eqref{psi_bounce} and cyclic \eqref{psi_ciclica} pilot-waves determined above, we apply Bohm-de Broglie quantum interpretation, by evaluating $\partial_jS_0$ and $\partial_jS_1$ for each solution $\psi$ and then inserting the results into \eqref{eq_guias_vel}, thus obtaining a dynamical system for each $\psi$.

By following this process, taking into account what was mentioned at the end of Section \ref{section_bounce}, we obtain two dynamical systems from \eqref{eq_guias_vel}, one for each pilot-wave, that will provide us the cosmological dynamics in the phase space $\varphi\times\alpha$. For the bounce solution \eqref{psi_bounce}, see Figure \ref{fig2} and for the cyclic one \eqref{psi_ciclica} see Figure \ref{fig3}. 

Now we can comment the results shown in these Figures. First of all, let us remember that the 0-th order solutions are \eqref{psi0nelson} (studied in \cite{PhysRevD.97.083517}), which we call the bounce solution, and \eqref{psi0_ciclica} (adapted from \cite{Torres_2020}), which we call the cyclic solution, because those are respectively the most common type of solutions for each system. These represent the behavior of the system without the perturbed term $-\kappa G^{\mu\nu}\partial_{\mu}\phi\partial_{\nu}\phi$, the nonminimal derivative coupling.

Thus, we can see the effect of the nonminimal coupling by the deviation from those 0-th order solutions in phase space. Hence, in Figures \ref{fig2} and \ref{fig3} we readly see that as $\alpha=\ln a$ decreases, the perturbations dominate. Conversely, when $\alpha$ increases, the general relativity terms dominate. Since $\kappa$ is the perturbative coefficient, it is expected that the smaller it is, the system will also approach the unperturbed cases, which are quantizations of general relativity. And this is consistent with the results also.

With these general properties in mind, we can analyse in more detail the results with respect to the questions that motivated our study, which are: the avoidance of initial singularity; the presence of accelerated expansion; the validity of gravitational waves speed constraint.

\subsection{Presence of Non-Singular Solutions}

As it is well known, a singularity problem arises in various scenarios in cosmology, and the quantization aims to avoid it, usually by replacing the classical singularity with a bouncing or a cyclic solution \cite{thebault_cqg_2023,struyve_nature_2017}. Even though we started from bouncing and cyclic solutions, one possibility is that the NDC perturbation disturb the system so much that singular solutions arise. In fact, as we can see in Figures \ref{fig2} and \ref{fig3}, that is not the case. In other words, the non-singular solutions present in the unperturbed case are modified, but they seem to remain  non-singular.

In Figure \ref{fig2}, we can also see cyclic solutions around center points, which occupy a smaller area in phase space than the bounce solutions. Since we are considering a Bohm-de Broglie interpretation, this means that if the theoretical universe will follow a bounce or a cyclic solution depends on the initial conditions $\alpha(0)$ and $\varphi(0)$. Here, the term \textit{initial} does not mean that the universe started at some initial time, since all solutions represent eternal universes. Rather, it is just the usual nomenclature form differential equations.

Even though Bohm-de Broglie is not probabilistic, it seems reasonable to think that any statistical analysis of the phase space has to lead with some probability measured defined over the phase space. In that point of view we could argue that, in some loose sense, bounces are more likely to happen than cyclic solutions. But since such a probabilistic analysis is an open question in quantum cosmology, we cannot assure that.

Besides bouncing and cyclic solutions, this systems also presents attractor and source critical points, which in closer look reveal solutions that are not bouncing or cyclic. Because of that, and also for the enormous analytical complexity of the dynamical system, we expect to investigate better the properties of those solutions that start in a source and evolve to an attractor in a future work.

A similar behavior occurs for the cyclic solutions shown in Figure \ref{fig3}, in the sense that the 0-th order cyclic solutions are modified but, for the region analysed, the universe still has a cyclic nature, without singularities. Figure \ref{fig3} (a) illustrates that attractors and a sources are present in the perturbation of the cycles too, which we also aim to investigate further in the aforementioned upcoming paper.

\subsection{Regions of Accelerated Expansion}

We can determine the regions in phase space where there is accelerated expansion (a.e.), as follows. Observe that $\alpha = \ln a$ implies $\ddot{a} > 0$ if, and only if, $\ddot{\alpha}+\dot{\alpha}^2 > 0$. Then, in the notation of \eqref{din_sys}, it follows that the acceleration condition is
\begin{equation}\label{condicao_de_aceleracao}
	f_{,\alpha}^{0}f^{0} + f_{,\varphi}^{0}g^{0}+(f^0)^2+\kappa\left( f_{,\alpha}^{0}f^{1} + f_{,\alpha}^{1}f^{0} + f_{,\varphi}^{0}g^{1} + f_{,\varphi}^{1}g^{0} + 2f^0f^1 \right) > 0\; ,
\end{equation}
up to first order in $\kappa$. To show that, we just used basic chain rule to express $\ddot{\alpha}$ in terms of $f^{i}$, $g^{i}$ and their derivatives.

In the standard cosmological model, the universe is always expanding, so $\ddot{a} > 0$ is the necessary and sufficient condition for a.e. But in bouncing and cyclic solutions, there are periods of time when the universe is contracting, so we must also suppose that $\dot{a}>0$, which is equivalent to $\dot{\alpha}>0\Leftrightarrow f^0+\kappa f^1>0$.

In both Figures \ref{fig2} and \ref{fig3}, we see the a.e. points shown as blue regions. For a given pair of initial conditions $\alpha(0)$, $\varphi(0)$ it is not unlikely that the solution will eventually experience some a.e. epoch. But to be precise one must select a particular solution, for some given initial conditions.

We can see that for the bounce solutions shown in Figure \ref{fig2} (a), a perturbation of order $\kappa = 10^{-2}$ has a wider region of points with a.e. than a smaller value of $\kappa$, like the ones shown in Figure \ref{fig2} (b) and (c). So this indicates that the perturbed term $-\kappa G^{\mu\nu}\partial_{\mu}\phi\partial_{\nu}\phi$ is causing that acceleration. When $\kappa$ it is smaller, a regions of acceleration gets much smaller than before, as we see in Figure \ref{fig2} (b) and (c), thus reducing the duration of acceleration for some solutions.

For the cyclic solutions shown in Figure \ref{fig3}, we see they are much more stable with respect to the a.e. regions, roughly for $\alpha\gtrsim 0$. For $\alpha\lesssim 0$, the perturbation terms dominate, so Figures \ref{fig3} (a), (b), and (c) show very different behaviors. In the $\alpha\lesssim 0$ region, with a domination of the perturbation term, we see the same property shown for the bouncing solution: the regions of a.e. decrease with $\kappa$, thus associating this term with a more likely accelerated expansion.

\subsection{Gravitational Waves Constraint}

As mentioned in the Introduction, we also aim to determine how the speed of gravitational waves behave in the quantized version of NDC theory. Now, since we are considering only the case $0<\kappa\ll 1 $, we must expand \eqref{cgw_geral} up to the first order in $\kappa$, resulting
\begin{equation}
	c_{gw} = 1-3\kappa\dot{\varphi}_0^2 \; ,	\label{cgw_perturb}
\end{equation}
where $\dot{\varphi}_0$ is the $g^{0}$ term from \eqref{din_sys}. In other words, the LIGO constraint in the perturbative quantum version of NDC is given by:
\begin{equation}\label{constraint_ligo_q}
	-3\times10^{-15} < -3\kappa\dot{\varphi}_0^2 \leq 7\times10^{-16}\;.
\end{equation}
The points $(\varphi,\alpha)$ that satisfy this constraint are shown in Figures \ref{fig2} and \ref{fig3} as gray regions, for some values of $\kappa$.

From these Figures, we can see that the smaller $\kappa$ gets, the wider the region compatible with the constraint is. This is expected, because the term $-3\kappa\dot{\varphi}_0^2$ is the deviation from unity, and it is proportional to $\kappa$. The superposition of blue and gray regions represent the points for which the constraint over $c_{gw}$ is valid and there is accelerated expansion at the same time.

For $\kappa = 10^{-2}$ it is easier to see that, for both bouncing and cyclic Figures, there are some pathological solutions near the boundary of the constrained region, that are compatible with the constraint for some period of time, and then evolve to a state where the constraint is no longer valid. 

But we can also see that, for $\alpha\gtrsim 6$, in all of our example plots, there are solutions that lie inside the constrained region, experience a period of accelerated expansion, and are non-singular solutions at the same time.

\section{Conclusion}\label{section_conc}
From the exposed above, we can say that it is possible to implement a canonical quantization to the theory, at least for a small value of the coupling parameter $\kappa$. This quantized version exhibits some differences from the classical theory (which we reviewed in \cite{ndc_class}) we would like to highlight. 

In the classical case, there are solutions of accelerated expansion asymptotically for $t\rightarrow -\infty$ that could, in principle, be used to build an inflationary theory. Some authors already cited pointed out that there would be a problem with the propagation of tensor perturbations, and we have shown in \cite{ndc_class} the precise relation between those two facts by studying the regions in phase space where there is acceleration and the region compatible with the gravitational waves, which is almost negligible. In this paper, we have implemented the same idea for the quantum version, looking for the superposition of the gravitational waves constraint with accelerated expansion points.

We have show above this superposition indeed happens in phase space, above some value for the scale factor, in Planck units. In this sense, the quantized version partially solves that incompatibility, so a period of primordial acceleration (alternative to inflation, as expected by the original work \cite{PhysRevD.80.103505} and its references) can be described without having to violate an already established constraint.

Roughly speaking, the smaller the scale factor is, the stronger the quantum effects are, so bellow this well behaved region the speed of gravitational waves become incompatible with the constraint again, as shown in Figures \ref{fig2} and \ref{fig3}. On the other hand, when the scale factor gets out of Planck scale, the classical behavior must come back, as the quantum potential decays. In other words, the compatibility we have shown is limited to the regime with a quantum effect, which must be not strong enough to break the constraint.

Another difference from the classical case is the meaning of the coupling constant $\kappa$. In the classical case, we were able to re-scale both $\alpha$ and $\varphi$ by a $\sqrt{\kappa}$ factor, so the qualitative properties of phase space is independent of $\kappa$. But the quantum version for small $\kappa$ changes significantly for different values of this parameter, and this effect increases when $\alpha$ decreases, as shown in the Figures \ref{fig2} and \ref{fig3}.

For future works, we expect to answer some questions. First, if it is possible to constrain $\kappa$ by the duration of the accelerated expansion, as it was done for the classical theory \cite{PhysRevD.80.103505}. Second, we would like to investigate other solutions for the Wheeler-DeWitt equation. Finally, we have to investigate further the attractor and source points of the guidance equations shown above, and what implications they may have for the evolution of the universe, in the quantum scale.

\section*{Acknowledgments}
We thank to Dilberto da Silva Almeida Junior, Felipe Tovar Falciano, Ingrid Ferreira da Costa, J\'ulio C\'esar Fabris, and Nelson Pinto-Neto for very important discussions about this paper. This study was financed in part by the \emph{Coordena\c{c}\~ao de Aperfei\c{c}oamento de Pessoal de N\'ivel Superior} - Brazil (CAPES) - Finance Code 001, and also by CNPq project No. 404310/2023-0 and FAPES, from Brazil.

\bibliographystyle{unsrt}
{\footnotesize
	\bibliography{ref_ndc_quant}

\begin{thebibliography}{10}

\bibitem{pdg_review_2022}
R.L.~Workman {\it et al.}~(Particle Data~Group).
\newblock {Review of Particle Physics}.
\newblock {\em Progress of Theoretical and Experimental Physics},
  2022(8):083C01, 08 2022.

\bibitem{turok_2014}
N.~Turok.
\newblock Tom kibble and the early universe as the ultimate high energy
  experiment.
\newblock {\em International Journal of Modern Physics A}, 29(06):1430015,
  2014.

\bibitem{borde_e_vilenkin_1996}
A.~Borde and A.~Vilenkin.
\newblock Singularities in inflationary cosmology: A review.
\newblock {\em International Journal of Modern Physics D}, 05(06):813--824,
  1996.

\bibitem{universe7120491}
L.~Fernández-Jambrina.
\newblock Singularities in inflationary cosmological models.
\newblock {\em Universe}, 7(12), 2021.

\bibitem{ndc_class}
{I. Torres, F.M. Santos}.
\newblock Nonminimal derivative coupling cosmology and the speed of
  gravitational waves, 2022.

\bibitem{PhysRevD.80.103505}
\relax S.V.~Sushkov.
\newblock {Exact Cosmological Solutions with Nonminimal Derivative Coupling}.
\newblock {\em Phys. Rev. D}, 80:103505, Nov 2009.

\bibitem{Kobayashi_2019}
T.~Kobayashi.
\newblock Horndeski theory and beyond: a review.
\newblock {\em Reports on Progress in Physics}, 82(8):086901, jul 2019.

\bibitem{HEISENBERG20191}
{L. Heisenberg}.
\newblock A systematic approach to generalisations of general relativity and
  their cosmological implications.
\newblock {\em Physics Reports}, 796:1 -- 113, 2019.
\newblock A systematic approach to generalisations of General Relativity and
  their cosmological implications.

\bibitem{nelson_2005}
{ N. Pinto-Neto}.
\newblock {{The Bohm Interpretation of Quantum Cosmology}}.
\newblock {\em Found. Phys.}, 35:577--603, 2005.

\bibitem{HOLLAND199395}
P.~Holland.
\newblock {The de Broglie-Bohm Theory of Motion and Quantum Field Theory}.
\newblock {\em Physics Reports}, 224(3):95 -- 150, 1993.

\bibitem{novello_2008_bouncing}
{M. Novello, S.E.P. Bergliaffa}.
\newblock Bouncing cosmologies.
\newblock {\em Physics Reports}, 463(4):127 -- 213, 2008.

\bibitem{steinhardt_prd_2002}
{P.J. Steinhardt, N. Turok}.
\newblock Cosmic evolution in a cyclic universe.
\newblock {\em Phys. Rev. D}, 65:126003, May 2002.

\bibitem{bojowald2011quantum}
M.~Bojowald.
\newblock {\em Quantum Cosmology: A Fundamental Description of the Universe}.
\newblock Lecture Notes in Physics. Springer New York, 2011.

\bibitem{PhysRevLett.119.161101}
{B.P. Abbott {\it et al.} (LIGO Scientific Collaboration and Virgo
  Collaboration)}.
\newblock {GW170817: Observation of Gravitational Waves from a Binary Neutron
  Star Inspiral}.
\newblock {\em Phys. Rev. Lett.}, 119:161101, Oct 2017.

\bibitem{Goldstein_2017}
{A. Goldstein {\it et al.}}
\newblock {An Ordinary Short Gamma-Ray Burst with Extraordinary Implications:
  Fermi-{GBM} Detection of {GRB} 170817A}.
\newblock {\em The Astrophysical Journal}, 848(2):L14, oct 2017.

\bibitem{Abbott_2017}
{B.P. Abbott {\it et al.}}
\newblock {Gravitational Waves and Gamma-Rays from a Binary Neutron Star
  Merger: GW170817 and GRB 170817A}.
\newblock {\em The Astrophysical Journal}, 848(2):L13, oct 2017.

\bibitem{lang2005algebra}
S.~Lang.
\newblock {\em Algebra}.
\newblock Graduate Texts in Mathematics. Springer New York, 2005.

\bibitem{acacio_nelson_problema_do_tempo_e_sing}
{J.A. de Barros, N. Pinto-Neto}.
\newblock {The Causal Interpretation of Quantum Mechanics and the Singularity
  Problem and Time Issue in Quantum Cosmology}.
\newblock {\em International Journal of Modern Physics D}, 07(02):201--213,
  1998.

\bibitem{nocedal2006numerical}
Jorge Nocedal and Stephen~J. Wright.
\newblock {\em Numerical Optimization}.
\newblock Springer, New York, 2006.

\bibitem{wolfram2019mathematica}
Inc. Wolfram~Research.
\newblock {\em Mathematica}, 2019.

\bibitem{PhysRevD.57.4707}
{R. Colistete Jr., J.C. Fabris, N. Pinto-Neto}.
\newblock Singularities and the classical limit in quantum cosmology with
  scalar fields.
\newblock {\em Phys. Rev. D}, 57:4707--4717, Apr 1998.

\bibitem{PINTONETO2000194}
{N. Pinto-Neto, A.F. Velasco, R. Colistete}.
\newblock {Quantum Isotropization of the Universe}.
\newblock {\em Physics Letters A}, 277(4):194 -- 204, 2000.

\bibitem{Colistete:2000ix}
{R. Colistete Jr., J.C. Fabris, N. Pinto-Neto}.
\newblock {Gaussian superpositions in scalar tensor quantum cosmological
  models}.
\newblock {\em Phys. Rev. D}, 62:083507, 2000.

\bibitem{PhysRevD.97.083517}
{A.P. Bacalhau, N. Pinto-Neto, S.D.P. Vitenti}.
\newblock Consistent {S}calar and {T}ensor {P}erturbation {P}ower {S}pectra in
  {S}ingle {F}luid {M}atter {B}ounce with {D}ark {E}nergy {E}ra.
\newblock {\em Phys. Rev. D}, 97:083517, Apr 2018.

\bibitem{Torres_2020}
{I.Torres, J.C. Fabris, O.F. Piattella}.
\newblock Bouncing and cyclic quantum primordial universes and the ordering
  problem.
\newblock {\em Classical and Quantum Gravity}, 37(10):105005, apr 2020.

\bibitem{thebault_cqg_2023}
K.P.Y. Thébault.
\newblock Big bang singularity resolution in quantum cosmology.
\newblock {\em Classical and Quantum Gravity}, 40(5):055007, feb 2023.

\bibitem{struyve_nature_2017}
W.~Struyve.
\newblock Loop quantum cosmology and singularities.
\newblock {\em Scientific Reports}, 7(1):8161, Aug 2017.

\end{thebibliography}
}

\end{document}